\pdfoutput=1
\documentclass[11pt]{article}

\usepackage[preprint]{acl}

\usepackage{times}
\usepackage{latexsym}

\usepackage[T1]{fontenc}
\usepackage[utf8]{inputenc}

\usepackage{microtype}

\usepackage{inconsolata}

\usepackage{graphicx}
\usepackage{booktabs}
\usepackage{array}
\usepackage{makecell}
\usepackage{siunitx}
\usepackage{multirow}
\usepackage{float}
\usepackage{placeins}

\usepackage{amsmath}
\usepackage{enumitem}
\setlist[itemize]{leftmargin=10pt, itemsep=0pt, topsep=0pt}
\setlist[enumerate]{leftmargin=15pt, itemsep=0pt, topsep=0pt}

\title{Knowledge Tracing in Programming Education \\ Integrating Students' Questions}

\author{Doyoun Kim \\
  Graduate School of Data Science \\
  Seoul National University \\
  South Korea \\
  \texttt{xxxdokki@snu.ac.kr} \\\And
  Suin Kim \\
  elice \\
  South Korea \\
  \texttt{suin.kim@elicer.com} \\ \\\And 
  Yohan Jo \\
  Graduate School of Data Science \\
  Seoul National University \\
  South Korea \\
  \texttt{yohan.jo@snu.ac.kr}}

\author{
 \textbf{Doyoun Kim\textsuperscript{1}},
 \textbf{Suin Kim\textsuperscript{2}},
 \textbf{Yohan Jo\textsuperscript{1}}
\\
 \textsuperscript{1}Seoul National University,
 \textsuperscript{2}Elice
\\
\texttt{\{xxxdokki, yohan.jo\}@snu.ac.kr}, \texttt{suin.kim@elicer.com}
\\
}

\begin{document}
\maketitle
\begin{abstract}
Knowledge tracing (KT) in programming education presents unique challenges due to the complexity of coding tasks and the diverse methods students use to solve problems. Although students' questions often contain valuable signals about their understanding and misconceptions, traditional KT models often neglect to incorporate these questions as inputs to address these challenges. This paper introduces SQKT (Students' Question-based Knowledge Tracing), a knowledge tracing model that leverages students' questions and automatically extracted skill information to enhance the accuracy of predicting students' performance on subsequent problems in programming education. Our method creates semantically rich embeddings that capture not only the surface-level content of the questions but also the student's mastery level and conceptual understanding. Experimental results demonstrate SQKT's superior performance in predicting student completion across various Python programming courses of differing difficulty levels. In in-domain experiments, SQKT achieved a 33.1\% absolute improvement in AUC compared to baseline models. The model also exhibited robust generalization capabilities in cross-domain settings, effectively addressing data scarcity issues in advanced programming courses. SQKT can be used to tailor educational content to individual learning needs and design adaptive learning systems in computer science education.\footnote{This paper is under review. We will release our source code upon the publication of the paper.}
\end{abstract}

\section{Introduction}
\begin{figure}[t]
    % \centering
    \includegraphics[width=0.47\textwidth]{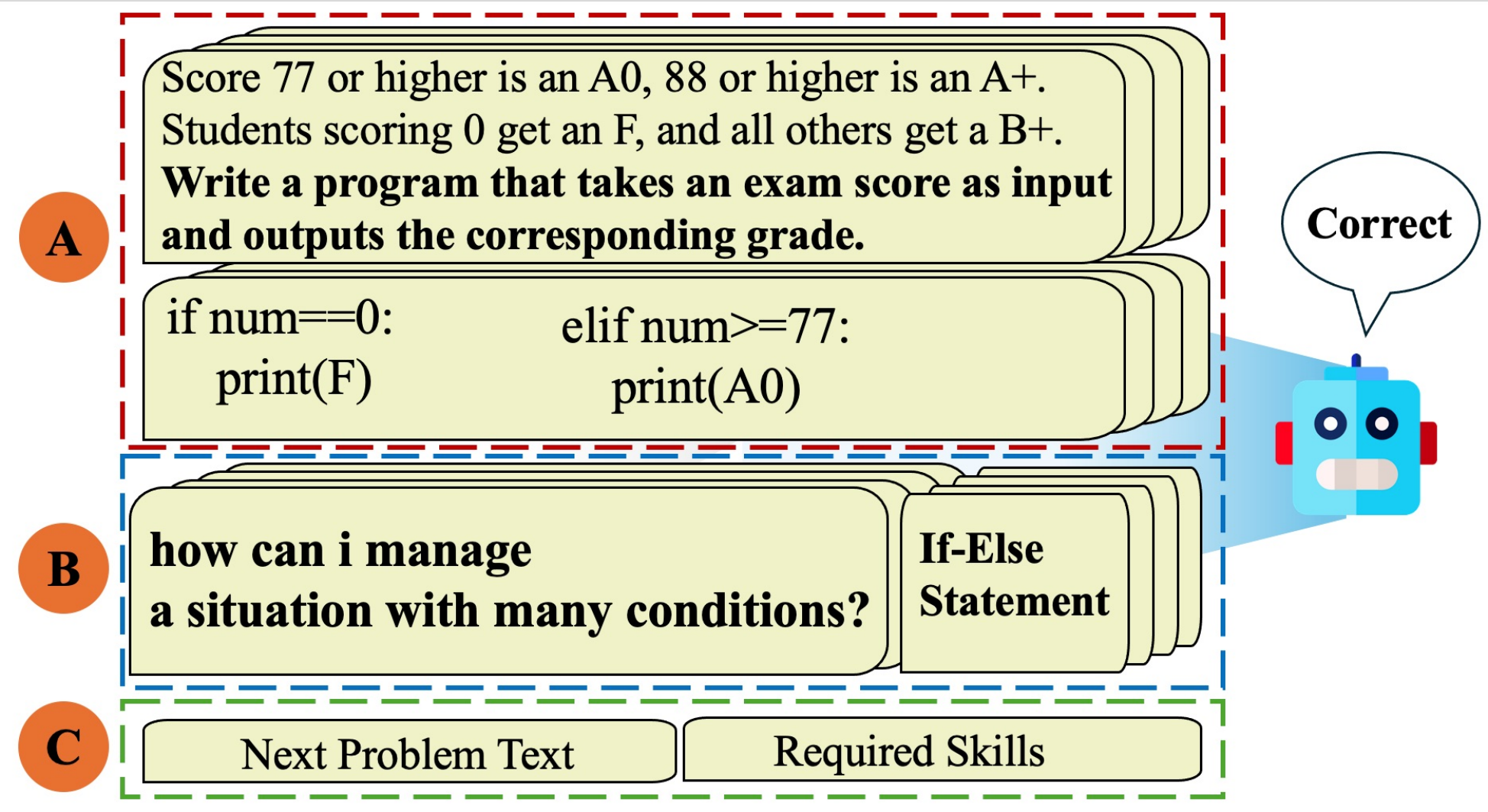} 
    % Reduce the figure size so that it is slightly narrower than the column.
    \caption{SQKT's process using an example from our dataset. A: All problem descriptions and code submissions from the student's history. B: The questions the student asked between submissions and the related skills extracted from these questions. C: The description of the next problem and the required skills inferred from the reference solution. The model uses the information from A and B and predicts the student's success or failure on the next problem.}
    \label{model_fig}
\end{figure}
Recent advancements in educational technologies have enabled the collection of dynamic data as students interact with learning systems. Consequently, researchers have paid considerable attention to knowledge tracing (KT), which involves monitoring students' knowledge states and predicting their future performance \cite{corbert1994knowledge}. 
A valuable source of signals about students' understanding and misconceptions is the questions they ask \cite{sun2021comparing}. With the growing popularity of online learning platforms and learning management systems (e.g., Moodle and Canvas) that include Q\&A forums, student questions and interactions with educators have become increasingly accessible. 
However, traditional KT models overlook this rich source of information. 
This gap is particularly significant in programming education, where KT is challenging because students' competencies need be assessed from unstructured and noisy source code. In such contexts, students' questions offer clearer insights into their understanding and confusion \cite{king1994guiding}. 

In this paper, we present the first model that integrates rich signals from student questions to accurately predict students' performance on subsequent problems, as illustrated in Figure \ref{model_fig}. As we will show, merely using a transformer to encode students' questions is suboptimal, because it might not fully represent the patterns of confusion and educational context that could be captured through the interaction between student and educator. Hence, our model enriches embeddings with two auxiliary signals: educator responses and skill information auto-extracted by GPT from students' questions. This approach creates a more comprehensive and detailed representation of the student's understanding, leading to improved prediction accuracy.

Experimental results show significant improvements over existing methods, with up to a 33.1\% absolute improvement in AUC compared to baseline models in in-domain experiments. Our model's ability to generalize across diverse educational content, including unseen courses with limited data, highlights its robustness. Our analysis reveals that this performance boost stems from the questions and the automatically extracted skill information, which offer insights into conceptual understanding and reasoning processes that are difficult to capture from code submissions alone. The combination of student questions with dynamically extracted skill information enables more accurate and granular modeling of student knowledge states. 
%This approach generalizes effectively to specialized programming courses with limited data.
% This work significantly advances knowledge tracing in programming education, offering a more comprehensive view of student learning beyond code submissions. 
Our approach is expected to contribute to more personalized and effective learning interventions in programming education.

The main contributions of this paper are:
\begin{itemize}
    \item To the best of our knowledge, this is the first work to integrate students' questions into KT, enabling more accurate predictions of students' success or failure on subsequent problems.
    \item  Our method for combining auto-extracted Python skills with student questions significantly improves model performance compared to relying solely on natural language questions.
    \item  Our model's strong performance in cross-domain settings highlights its generalizability across different course materials and environments.
\end{itemize}

\section{Related Works}

\paragraph{Knowledge Tracing with Behavioral Data} Knowledge tracing (KT) models students' knowledge over time to predict their future performance \cite{piech2015deep}. Building upon the foundational approaches like Bayesian Knowledge Tracing (BKT) \cite{corbert1994knowledge} and Deep Knowledge Tracing (DKT) \cite{piech2015deep}, recent research has advanced KT by incorporating behavioral data, such as response times \cite{song2021deep}, scaffolding interactions \cite{asselman2020evaluating}, and attempt counts \cite{sun2022dynamic}.

However, a significant gap remains in leveraging student-educator interactions. Questions arising from these interactions often reveal students' reasoning processes and areas of struggle in applying theoretical knowledge to coding \cite{sun2021comparing}. Yet, most existing models fail to capture the valuable insights embedded in students' questions. Our model addresses this challenge by directly integrating this rich data.

% Our framework extends these methods by integrating  both students' code submissions and their related questions that they ask before writing the code. We use these inputs to enhance model's predictive accuracy. By analyzing these inputs, we can capture not only the outcomes of their attempts but also the implicit knowledge and areas of confusion reflected in their questions.

\paragraph{Knowledge Tracing in Programming Education} Programming education poses unique challenges for KT due to the complexity of coding tasks and multiple correct solutions that can be derived using various skills. Traditional KT models often use the Q-matrix method to manually tag problems with the required skills \cite{help-dkt}, but this process is labor-intensive and often fails to capture the full range of skills students use. The diversity in problem-solving approaches complicates the tracing of specific skills mastered by a student, making it challenging to predict future performance accurately.

A key aspect of KT in programming education is the representation and modeling of knowledge components (KCs), such as ``for loop'', ``recursion'', or ``object-oriented principles''.  Recent work has focused on analyzing code submissions to model these KCs and predict learning states. \citet{price2022code} introduced Code-DKT, which uses attention mechanisms to extract domain-specific code features. \citet{liu2022open} developed an approach that considers the multi-skill nature of programming exercises by learning features from student code that reflect multiple skills.

However, these approaches still rely on manual tagging of KCs. Our approach advances this by using an automated skill-mapping system using GPT to extract KCs from student questions. This method allows more flexible use of KCs, enabling the model to identify and leverage skills without extensive manual tagging. This skill extraction method captures aspects of student knowledge that are not evident in code submissions alone, improving the prediction of student performance.

\section{Methods}

\begin{figure*}[t]
\centering
\includegraphics[width=0.8\textwidth]{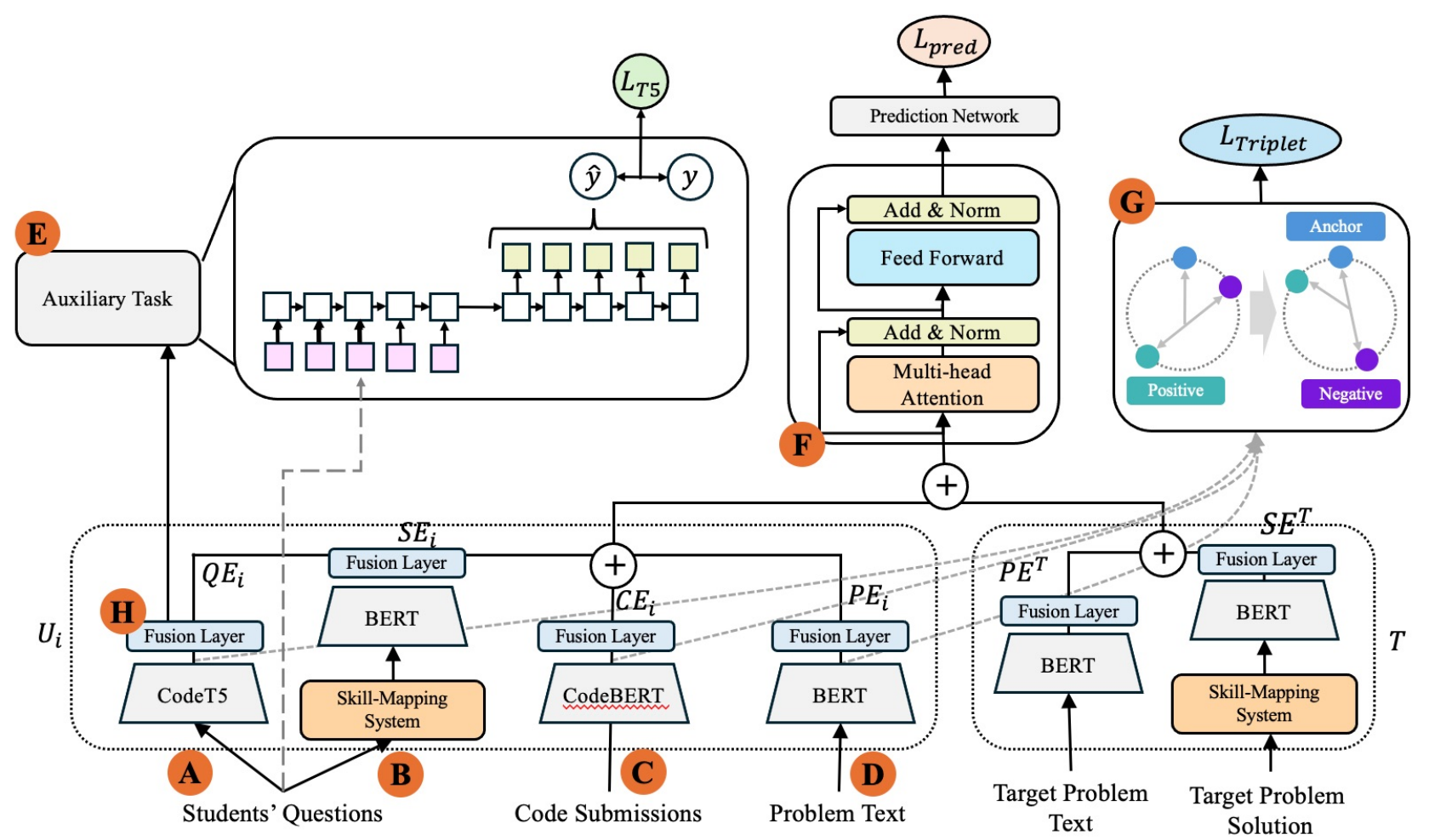}
\caption{Comprehensive architecture of the SQKT. The model processes problem text, code submissions, and student questions through three embedding layers. Skill extraction is performed using a GPT-based skill-mapping system. All embeddings and extracted skills are combined through a fusion layer, which is then processed by transformer encoder layers to generate the final prediction output. The model is trained using multiple objective functions, including \(L_{triplet}\) for aligning the diverse embeddings and \(L_{pred}\) for predicting students' performances on tasks. Additionally, the 
auxiliary objective function \(L_{question}\) is included to enhance the model's robustness and generalization capabilities.}
\label{SQKT_fig}
\end{figure*}

In this section, we introduce our Students' Question-based Knowledge Tracing (SQKT) model. 
Our primary goal is to predict a student's success on a problem by integrating information about the student's history of solving other problems. Our model takes a sequence of descriptions of problems the student has attempted in the past, associated code submissions, student questions, and skill information (each problem may have multiple submissions and questions), along with the description and required skills for the next problem. 
The model then predicts whether the student will correctly solve the next problem. The overall architecture is illustrated in Figure~\ref{SQKT_fig}.

\subsection{Multi-feature Inputs} 

\begin{figure}[t]
\centering
\includegraphics[width=0.4\textwidth]{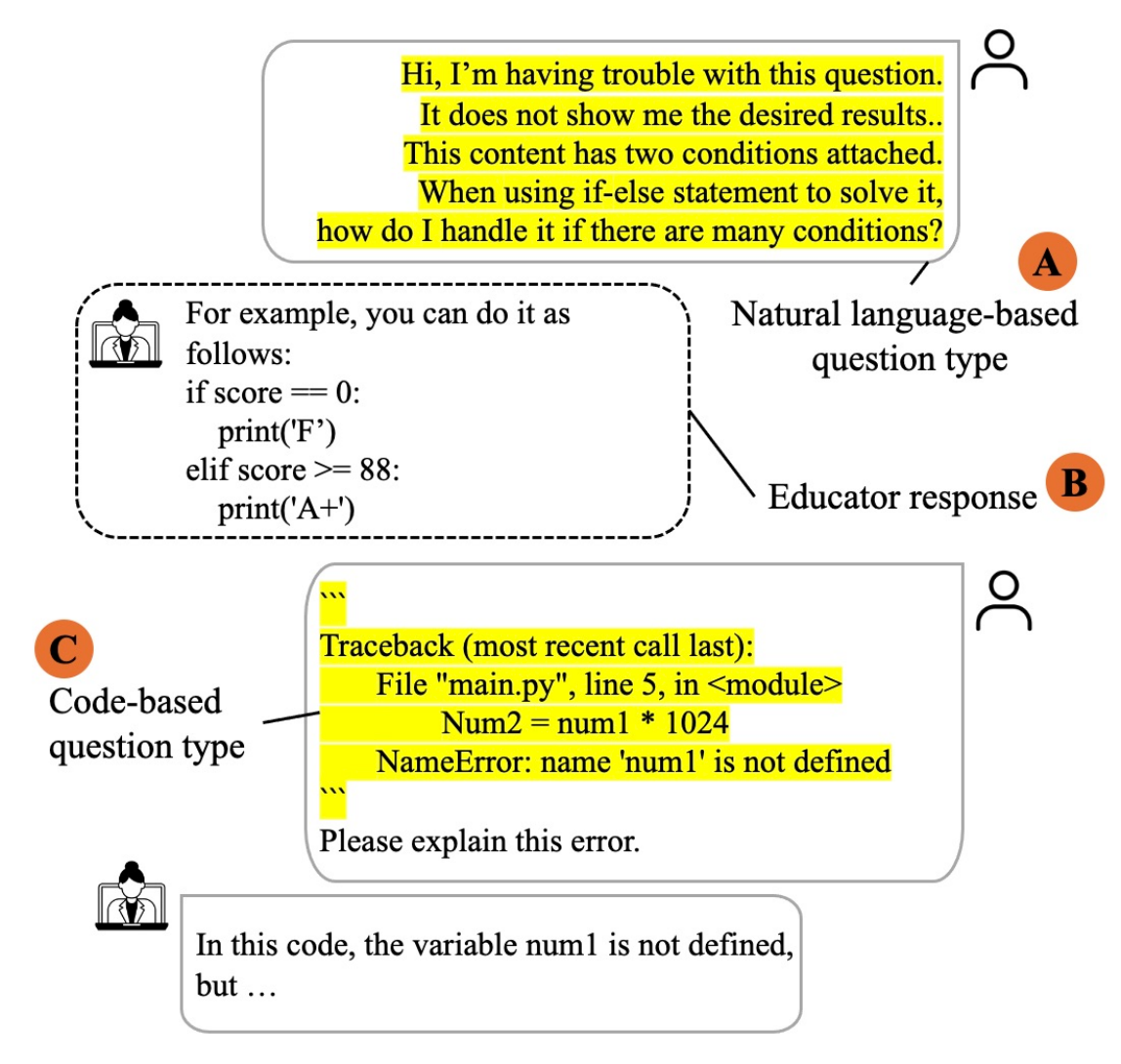} % Reduce the figure size so that it is slightly narrower than the column.
\caption{This figure illustrates different types of student questions and interactions. (A) Natural language-based questions (B) Educator responses (C) Code-based questions}
\label{SQ_fig}
\end{figure}

SQKT integrates various input features, with a focus on students' questions and extracted skills. In this section, we first detail our main contribution: the integration of students' questions as input features, followed by an overview of the remaining input components.

\paragraph{Student Questions (Figure \ref{SQKT_fig}, A)} Integrating student questions is motivated by valuable insights they provide into a student's mastery level, revealing areas of confusion and the depth of understanding of specific concepts \cite{sun2021comparing}.
As illustrated by an example student-educator interaction in Figure~\ref{SQ_fig}, students' questions typically include two types of information: natural language questions that seek clarification on specific concepts or strategies (A), and code-based questions that address specific lines of code or errors (C). Educator responses further clarify misconceptions, provide additional context, and highlight key concepts (B). 

To effectively leverage this information, we employ the CodeT5 model \cite{codet5} for embedding student questions. CodeT5 was chosen for its ability to understand both natural language and code syntax, making it ideal for processing the mixed content of students' questions. If no student questions exist  when SQKT expects a question embedding, a zero vector is used instead.

We further enhance this embedding process by fine-tuning CodeT5 with an auxiliary task of generating potential educator responses (Figure \ref{SQKT_fig}, E). This task helps the question embedding capture the gist of a student's question (mainly confusion and erroneous code) that is predictive of the educator's response. 
The following auxiliary loss function is integrated into our overall training objective:
\begin{align}
L_{question} = -\Sigma_{(x,y)} \log P(y | x) \label{eq:L_T5}
\end{align}
where $(x, y)$ is a pair of student question and educator response. This enhances question embeddings and the model's overall prediction accuracy.

% \begin{table*}[t]
% \centering
% \small
% % \scriptsize
% \setlength{\tabcolsep}{2pt}
% \begin{tabular}{>{\raggedright\arraybackslash}p{0.18\linewidth} >{\raggedright\arraybackslash}p{0.18\linewidth} >{\raggedright\arraybackslash}p{0.18\linewidth} >{\raggedright\arraybackslash}p{0.18\linewidth} >{\raggedright\arraybackslash}p{0.18\linewidth}}
% \toprule
% Value & Variable Assign & Keywords & Operators & Operands \\
% Type Convertor & input function & print function & Boolean Values & Boolean Expressions \\
% Logical Operators & If-Else Statements & For Loops & While Loops & Break Statement \\
% Continue Statement & Function Definitions & return Statement & Local, Global Scope & Strings \\
% String Slicing & Indexing & Lists & Dictionaries & Import Statement \\
% random & time & math & Opening files & Reading files \\
% Writing files & Closing files & SyntaxError & NameError & TypeError \\
% IndentationError & ValueError & AttributeError & IndexError & KeyError \\
% TabError & UnicodeDecodeError & FileNotFoundError & ModuleNotFoundError & ZeroDivisionError \\
% UnboundLocalError & ImportError & UnicodeEncodeError & LookupError & ConnectionError \\
% RuntimeError &  &  &  &  \\
% \bottomrule
% \end{tabular}
% \caption{Categorized Python concepts and errors}
% \label{python_concepts_errors}
% \end{table*}

\paragraph{Skill Extraction (Figure \ref{SQKT_fig}, B)} 
Identifying the skills students struggle with can improve our model's performance compared to relying solely on questions. Extracting skills from student questions is more straightforward and accurate than from submitted codes, as these questions often directly address the concepts students find challenging. By combining these extracted skills with those required for the target problem, the model can predict a student's performance more accurately.
To achieve this, we developed a method to extract and leverage skill information from both student questions and target problems. 

The first challenge was to define an effective set of Python skills. We identified a comprehensive set of 36 core Python concepts and 19 Python error types, drawing from Python's official documentation and books by \citet{python:1} and \citet{python:2}, as shown in Table \ref{python_concepts_errors}. Incorporating error types as skills was motivated by the pedagogical principle that errors reveal students' understanding and misconceptions, which are correlated with learning gaps \cite{altadmri201537, becker2019compiler, hertz2013investigating}. 

The next challenge was scaling skill extraction. Traditional approaches rely on experts to manually tag skills for each problem, which is labor-intensive and lacks scalability. To address this, we developed an automatic method using GPT-4o. Specifically, we provided GPT-4o with about 20 examples of student questions and a pre-defined list of skills. GPT-4o was then prompted to reference these examples and generate a Python script that could be used to map any student question to the relevant skills from our predefined skill list. The resulting skill extraction script, referred to as the \textit{skill extractor}, uses specific rules to identify skills from both natural text and code. We found that a rule-based method is preferable to using GPT on the fly, due to high precision and consistency in skill identification. 
We reviewed the script and corrected inaccurate or unreliable rules based on some student questions manually labeled with skills.
%To validate the skill extractor, we applied it to sample data and manually compared the output skills with the actual skills required in the questions. Based on this, we identified and corrected inaccurate or unreliable rules in the extractor. 

To validate the skill extractor more systematically, we evaluated it on a random sample of 100 student questions from the ``Python Basic'' course (\S{\ref{sec:dataset}}). These questions were annotated with ground-truth skills by a co-author of this study, and these annotations were further validated by a graduate student proficient in Python but not involved in this study, resulting in Cohen's kappa of 0.98. The skill extractor achieved a precision of 0.85, a recall of 0.88, and an F1-score of 0.86. These results indicate that the skill extractor produces reliable outputs that closely match human judgments. We considered this level of accuracy acceptable, as extracted skills substantially improve SQKT's predictive performance (as discussed in the experiment section).

We first use the skill extractor to extract skill information from student questions. Specifically, it processes student questions to identify the particular skills students are struggling with. These identified skills are concatenated as a single text and encoded into a \textit{skill embedding} using the pre-trained BERT-base model \cite{bert}. 
In addition, the skills required to solve each problem are identified by applying the skill extractor to the reference solution code provided with each problem in our dataset. 
Taken together, this approach enables SQKT to align the extracted skills with specific skills required for the target problem, thereby enhancing its predictive accuracy.

\paragraph{Code Embedding (Figure \ref{SQKT_fig}, C)} We use CodeBERT \cite{codebert}, a pre-trained transformer model designed for programming languages, to convert students' code submissions into vector representations. This captures both the syntactic and semantic properties of the code, providing insights into the student's coding abilities and problem-solving strategies.

\paragraph{Problem Embedding (Figure \ref{SQKT_fig}, D)} Problem descriptions include the problem statement, input/output specifications, and constraints. They are processed through the pre-trained BERT-base model \cite{bert} to generate an embedding that captures the contextual meanings of the problem statements. This information is crucial for understanding the task requirements and difficulty levels.

\paragraph{Fusion Layer (Figure \ref{SQKT_fig}, H)} The fusion layer combines the above embeddings---questions, skills, problem descriptions, and code submissions---into a unified representation space. While each source provides unique insights, challenges remain in integrating these heterogeneous signals. The fusion layer addresses this by projecting each embedding type into a common 512-dimensional space based on the relationships among the embeddings. 

Specifically, we employ triplet loss (Figure \ref{SQKT_fig}, G) to encourage embeddings from the same submission to be positioned closely together, while those from different submissions or representing distinct programming concepts are placed farther apart. The triplet loss is defined as:
\begin{align}
    L_{triplet} = max(0, d(a,p) - d(a,n) + margin), \label{eq:L_triplet}
\end{align}
where:
\begin{itemize}
    \item \(a\) is the current problem's embedding derived from the student's code submission, serving as the anchor embedding.
    \item \(p\) is the embedding of the current problem's description or student questions, serving as positive samples.
    \item \(n\) is the embedding of a randomly selected problem's description or student questions, serving as negative samples.
    \item \(d(x,y)\) is the Euclidean distance between two embeddings $x$ and $y$.
    \item \(margin\) is a hyperparameter enforcing a minimum distance between positives and negatives.
\end{itemize}
Consequently, the fusion layer enhances SQKT's ability to process heterogeneous yet semantically and contextually related signals more coherently.
% \begin{table*}[t]
% \centering
% \small
% \begin{tabular}{l r r p{9.5cm}}
% \toprule
% Course Name & \makecell{\# of Code \\ Submissions} & \makecell{\# of \\Interaction} & Description \\
% \midrule
% Python Basic& 3,246 & 1,375 & Beginner-level course covering basic python concepts. \\
% First Python & 167,453 & 127,439 & A beginner-friendly course focusing on fundamental Python programming. \\
% Algorithm & 453 & 976 & An intermediate to advanced course on algorithms implemented in Python. \\
% \makecell[l]{Python \\Introduction} & 58,580 & 42,568 & A comprehensive course covering basic to intermediate python concepts. \\
% \bottomrule
% \end{tabular}
% \caption{Summary and key statistics of dataset}
% \label{table:data}
% \end{table*}

\subsection{Multi-Head Self-Attention Layers} 
All embeddings from the student's history and the next problem are encoded through a multi-head attention mechanism to predict the student's success or failure on the next problem (Figure \ref{SQKT_fig}, F). 
The target problem for prediction is represented by the following tensor:
\[
    T = [PE^T, SE^T] \in R^{2 \times 512}
\]
where $PE^T$ and $SE^T$ are the problem and skill embeddings for the target problem.

For each problem \(i\) that the student attempted prior to the target problem, we construct a tensor \(U_i\) containing the input features associated with the $i$th problem:
\[
    U_i = [PE_i, CE_i, QE_i, SE_i] \in R^{K \times 512}
\]
where $PE_i$, $CE_i$, $QE_i$, and $SE_i$ denote the problem, code, student question, and skill embeddings, respectively. Each $CE_i$, $QE_i$, and $SE_i$ is a tensor with potentially multiple rows, consisting of embeddings accumulated from all code submissions and questions related to the $i$th problem. If the student asked no questions, $QE_i$ is set to a zero vector. $K$ increase as the student makes more submissions for the $i$th problem.

Taken together, the input to the multi-head self-attention layers consists of the target problem along with all preceding learning history $[U_1, U_2, \ldots, U_n , T]$. 

% \begin{itemize}
%     \item $\oplus$: vector concatenation. 
%     \item \(K\): the cumulative size of the embeddings, accounting for all combined features.
%     \item \(TE_i\): the \(i\)th problem description embedding.
%     \item \(CE_i\): the \(i\)th code submission embedding.
%     \item \(QE_i\): the embedding of the student's question before the \(i\)th submission, if any. If no question was asked, we replace this with a zero vector.
%     \item \(SE_i\): the \(i\)th identified skill embedding from the student's question.
%     \item \(TE_{i+1}\): the target problem description embedding.
%     \item \(SE_{i+1}\): the skill embedding from the solution code of the target problem.
% \end{itemize}

This input sequence passes through six self-attention layers, each capturing complex interactions among different submissions and their components. After the final attention layer, max-pooling is applied to all output embeddings to derive a representation of the student's knowledge state. The resulting embedding is then fed to a classification head to predict the student's success or failure on the target problem. Binary cross-entropy is used as the loss function:
\[
L_{pred} = -\Sigma (y \log(\hat{y}) + (1-y) \log(1-\hat{y})),
\]
where \(y\) is the true label and \(\hat{y}\) is the predicted label.

The final loss function is a weighted sum of the prediction loss, question loss (Eq.~\ref{eq:L_T5}) and triplet loss (Eq.~\ref{eq:L_triplet}):
\[
L_{total} = L_{pred} + L_{question} + \lambda L_{triplet}
\]
where \(\lambda\) is a hyperparameter that adjusts the weight of the triplet loss. 
% The model's goal is to minimize \(L_{total}\), ensuring accurately predicting the student's performance.

\section{Experiment Settings}
To evaluate the performance and generalizability of our SQKT model, we conduct experiments aimed at answering the following research questions:
\begin{enumerate}
    \item How does SQKT compare to existing knowledge tracing models in predicting student performance on programming problems?
    \item To what extent does the integration of question data and skill information enhance the model's predictive accuracy?
    \item How well does SQKT generalize across different courses with different difficulty levels?
\end{enumerate}

\subsection{Dataset}\label{sec:dataset}

% \begin{table}
% \centering
% \small
% \begin{tabular}{l r r}
% \toprule
% Course Name & \makecell{\# of Problems} & \makecell{\# of Interaction}\\
% \midrule
% Python Basic& 3,439 & 1,375\\
% First Python & 183,457 & 127,439\\
% Algorithm & 330 & 976\\
% \makecell[l]{Python Introduction} & 47,197 & 42,568\\
% \bottomrule
% \end{tabular}
% \caption{Summary and key statistics of dataset}
% \label{table:data}
% \end{table}

\begin{table}
\centering
\small
% \begin{tabular}{p{2.1cm} p{0.8cm} p{1cm} p{0.8cm} p{1cm}}
\begin{tabular}{%
    @{}
    p{1.6cm}
    >{\raggedleft\arraybackslash}p{0.9cm}
    >{\raggedleft\arraybackslash}p{1.3cm}
    >{\raggedleft\arraybackslash}p{0.8cm}
    >{\raggedleft\arraybackslash}p{1cm}}
\toprule
Attribute & PB & FP & Algo. & PI \\
\midrule
\makecell[l]{Unique\\problems} & 48 & 60 & 32 & 227\\
\makecell[l]{Submissions\\per problem} & 474 & 20,573 & 297 & 1,533\\
Students & 160 & 8,141 & 77 & 1,092 \\
Training & 17,685& 1,050,360 & 5,689 & 308,825\\
Validation & 2,161 & 123,975 & 2,233 & 20,991\\
Test & 2,926 & 60,071 & 1,587 & 18,251\\
\bottomrule
\end{tabular}
\caption{Statistics of the dataset. PB: Python Basic, FP: First Python, Algo: Algorithm, PI: Python Introduction.}
\label{table:data}
\end{table}

Our study uses data collected from a Korean online programming education platform between January 2022 and April 2024, with the consent of the copyright holders. These data cover four distinct Python programming courses, providing a diverse range of difficulty levels and topics. All data are in Korean and include Python code, covering code blocks and associated error messages. Statistics are summarized in Table~\ref{table:data} and description and examples are in Appendix~\ref{appendix:dataset} and ~\ref{appendix:example}. 
% Each course dataset consists of students' submissions, along with submission ID, creation time, problem ID, submitted code, and execution result. 
Additionally, the data contain student-educator interactions including student questions and educator answers (Figure \ref{SQ_fig}). 
% Each interaction includes a question/answer ID, creation time, content, and related problem ID. 
The data for each course are split by students into training, validation, and test sets in an 8:1:1 ratio, with each student assigned to only one set to prevent the risk of information leakage.

\subsection{Experimental Setup}
We conduct a series of experiments to assess two critical aspects: the model's ability to predict student performance and generalize across different courses and difficulty levels. We perform both in-domain and cross-domain experiments.

\paragraph{In-domain} We evaluate the model's performance when trained and tested on the same course. We experiment with three out of four courses, excluding one due to insufficient data for stable training. 

\paragraph{Cross-Domain} 
We selected courses to challenge the model's adaptability and generalization capabilities. In the first cross-domain setting, labeled `content structure generalization', we train the model on the ``Python Introduction'' course (5,858 samples) and tested it on the ``First Python'' course (1,674 samples). This pair was chosen to evaluate the model's ability to transfer knowledge between courses with different content structures, including varying difficulty levels and vocabulary usage. 

In the second cross-domain experiment, labeled `data-scarce generalization', we train the model on combined data from all courses except ``Algorithm'' (9,390 samples) and tested it exclusively on the ``Algorithm'' course (300 samples). This choice was motivated by our initial observation that the model struggled on this course due to the small data size. Through this setting, we aimed to verify the model's ability to generalize to a specialized, data-scarce course by leveraging student questions.

\paragraph{Baseline Models} 
To benchmark SQKT's performance and evaluate the impact of integrating student questions, we compare it with several baseline models. 
Our choice of baseline models is limited by the scarcity of prior KT models capable of processing code submissions without relying on predefined skill annotations. Moreover, to the best of our knowledge, no existing KT models incorporate student questions.
To that end, we experiment with four baseline models: KTMFF \cite{xiao2023knowledge}, OKT \cite{liu2022open}, and their variants. KTMFF and OKT are known for their strong performance in leveraging rich embeddings of code submissions and capturing the structural properties of code blocks. To verify the effectiveness our question embeddings and demonstrate their adaptability for enhancing different models, we introduce KTMFF+ and OKT+, variants of KTMFF and OKT that incorporate question embeddings as additional input. 
% This setup isolates the effect of question data on model performance.
%This allowed us to compare SQKT against leading approaches that emphasize the richness of code embeddings without directly incorporating question data.

% We selected these comparisons due to the limited availability of existing models that fully integrate raw textual data in knowledge tracing for programming education. Most current approaches rely on pre-defined labels, processing textual data to extract these labels before modeling. In contrast, our model and the baselines operate directly on full text data, enabling the models to learn from the richness and complexity of the raw text without being constrained by pre-defined labels.

\textbf{Evaluation Metrics} 
To evaluate each model's performance in predicting student success on programming problems, we use AUC, accuracy, and F1-score based on the model's predictions. Here, a student is considered successful on a problem if they achieve a score of 100 within a certain number of trials. This threshold is set to the average number of submissions across all students in each course. Any score below 100 or a submission count exceeding this threshold is considered a failure.

% three key metrics:    
% \begin{itemize}
% \item AUC: Measures the model's ability to distinguish whether the student will succeed or fail on the next problem.
% \item Accuracy: The proportion of correct predictions in determining whether a student will solve the next problem.
% \item F1-Score: The harmonic mean of precision and recall, balancing the model's ability to identify both student success and failure.
% \end{itemize}

\subsection{Training Setup} 
For each student, we predict the student's outcome for every problem they attempted, excluding the first problem since it has no preceding history. For any target problem, all the history $U$ preceding that problem is used as the model input.
Note that there is no risk of a student's history being exposed to testing, as students do not overlap between the training and test sets (see Table ~\ref{table:split_data}).

% Each instance comprises a question description, the student's code submissions for that question, accompanying questions (or a dummy embedding if none are present), skill information, and the target question's description, represented as a composite vector \(U\). Each mini-batch consists of a sequence of this information. 

% To ensure no overlap between data subsets, we assigned each student exclusively to either the training, validation, or test set. 
We conducted a grid search across a range of hyperparameters, including dropout rate, learning rate, batch size, and the weight for the triplet loss. 
The optimal hyperparameter values were chosen based on performance on the validation set. 
The best configuration obtained is as follows: a dropout rate of 0.1, a learning rate of 3e-5, a batch size of 16, and an auxiliary loss weight of 1.0.

The model is trained using the Adam optimizer on an NVIDIA A100 80GB PCIe GPU. The training times vary depending on the scenario: approximately 1 hour and 30 minutes for in-domain tasks and around 3 hours for cross-domain tasks. 

\section{Experiment Results}
\subsection{In-Domain Results}
% \begin{table}[t]
% \centering
% \small
% \setlength{\tabcolsep}{3.5pt}
% % \begin{tabular}{ll S[table-format=1.4] S[table-format=1.4] S[table-format=1.4]}
% \begin{tabular}{ll c c c}
% \toprule
% Course & Metric(\%) & KTMFF& KTMFF+& SQKT \\
% \midrule
% \multirow{3}{*}{Python Introduction} & AUC & 70.3 & 74.7 & \bfseries 92.0 \\
%                                      & ACC & 62.5 & 66.5 & \bfseries 90.6 \\
%                                      & F1  & 48.3 & 57.3 & \bfseries 89.2 \\
% \addlinespace
% \multirow{3}{*}{First Python} & AUC & 75.2 & 71.1 & \bfseries 89.6 \\
%                               & ACC & 68.2 & 62.3 & \bfseries 84.3 \\
%                               & F1  & 68.3 & 63.6 & \bfseries 83.9 \\
% \addlinespace
% \multirow{3}{*}{Python Basic} & AUC & 79.3 & 80.9 & \bfseries 87.7 \\
%                               & ACC & 74.0 & 74.8 & \bfseries 85.6 \\
%                               & F1  & 80.5 & 80.9 & \bfseries 88.2 \\
% \bottomrule
% \end{tabular}
% \caption{Performance comparison of various models across three datasets}
% \label{in-domain}
% \end{table}]

\begin{table}[t]
\centering
\small
\setlength{\tabcolsep}{1.2pt}
\begin{tabular}{l c c c c c c c c c}
\toprule
 & \multicolumn{3}{c}{\makecell{Python \\Introduction}} & \multicolumn{3}{c}{First Python} & \multicolumn{3}{c}{Python Basic} \\
\cmidrule(lr){2-4} \cmidrule(lr){5-7} \cmidrule(lr){8-10}
& AUC & ACC & F1 & AUC & ACC & F1 & AUC & ACC & F1 \\
\midrule
KTMFF  & 70.2 & 64.5 & 56.5 & 69.4 & 61.8 & 60.3 & 78.0 & 73.5 & 80.0 \\
KTMFF+ & 72.6 & 66.1 & 58.2 & 71.7 & 62.1 & 60.7 & 80.7 & 76.1 & 81.4 \\
OKT & 60.3 & 81.0 & 34.6 & 65.8 & 77.7 & 49.1 & 65.0 & 78.8 & 46.2\\
OKT+ & 66.7 & 83.3 & 49.8 & 66.7 & 83.3 & 49.8 & 78.4 & 82.1 & 70.2 \\
SQKT   & \bfseries 93.4 & \bfseries 89.2 & \bfseries 88.4 & \bfseries 90.3 & \bfseries 87.1 & \bfseries 84.9 & \bfseries 93.3 & \bfseries 88.4 & \bfseries 89.8 \\
\bottomrule
\end{tabular}
\caption{Performance comparison of various models across three datasets. All values are in percentages.}
\label{in-domain}
\end{table}
Across the three courses, SQKT consistently outperformed all baselines. 
SQKT achieved an AUC of 87.1--93.4, representing an absolute improvement of 12.6--20.8 compared to the best-performing baseline (KTMFF+).
These results demonstrate that our SQKT model, which incorporates student questions, is highly effective in predicting students' future performance.

% For the ``Python Introduction'' course (col 1--3), SQKT achieved an AUC of 93.4\%, improving by 20.8\% over KTMFF+ and 26.7\% over OKT+. The ACC and F1-scores also showed significant gains. For the ``First Python'' course (col 4--6), SQKT achieved an AUC of 90.3\%, outperforming KTMFF+ by 18.6\% and OKT+ by 23.6\%. ACC and F1-scores follow a similar trend. For the ``Python Basic'' course (col 7--9), SQKT achieved an AUC of 93.3\%, improving by 12.6\% over KTMFF+ and 14.9\% over OKT+. Similarly, ACC and F1-scores showed improvements.

The improvement of KTMFF+ over KTMFF and OKT+ over OKT reinforces our research motivation that student questions provide valuable insights into student performance. It also suggests that our question embeddings can be integrated with general KT models to enhance their predictive accuracy.
However, SQKT consistently outperformed these models, underscoring the efficacy of its architecture in leveraging student questions more effectively than the baselines.

\paragraph{Ablation Study}
\begin{table}[t]
\centering
\small
\setlength{\tabcolsep}{1pt} % Adjust column separation
\begin{tabular}{l S[table-format=2.1] S[table-format=2.1] S[table-format=2.1]}
% \begin{tabular}{l c c c}
\toprule
Model & {AUC (\%)} & {ACC (\%)} & {F1 (\%)} \\
\midrule
SQKT & \textbf{93.4} &  \textbf{89.2} &  88.4 \\
% - contents (Dummy question - Zero) & 89.1 & 84.7 & 83.2 \\
- Question (all-ones vector) &  91.3 & 86.3 & \textbf{89.9} \\ % one-vector
- Question (skill only) & 90.9 & 86.2 & 88.7 \\
- Skill (question only) & 89.7 & 81.3 & 83.1 \\
- Question and skill & 85.4 & 80.7 & 82.7 \\
\bottomrule
\end{tabular}
\caption{Ablation study on the ``Python Introduction'' course.}
\label{table:ablation_study}
\end{table}

% \begin{table}[t]
% \centering
% \small
% \setlength{\tabcolsep}{3pt} % Adjust column separation
% \begin{tabular}{l S[table-format=2.1] S[table-format=2.1] S[table-format=2.1]}
% \toprule
% Model & {AUC (\%)} & {ACC (\%)} & {F1 (\%)} \\
% \midrule
% SQKT & \bfseries 87.7 & \bfseries 85.6 & \bfseries 88.2 \\
% - contents (Dummy question) & 83.4 & 81.7 & 85.3 \\
% - question (Skill only) & 84.6 & 81.2 & 85.4 \\
% - skill (Question only) & 83.9 & 82.1 & 86.0 \\
% - question and skill & 83.7 & 77.1 & 82.4 \\
% \bottomrule
% \end{tabular}
% \caption{Ablation study on ``Python Basic'' dataset}
% \label{table:ablation_study}
% \end{table}
To evaluate the contribution of each component in SQKT, we conducted an ablation study.
Basically, we explore removing question embeddings and skill embeddings both separately and together to assess their impact. 
Additionally, to examine the importance of the actual content of student questions, we replace the question embeddings with an all-ones vector to simply indicate the presence of a student question (potentially student confusion).

Table \ref{table:ablation_study} presents the ablation results on the ``Python Basic'' course (the same pattern is observed in other courses). Using question indicators (row 2) reduces AUC and ACC, highlighting the importance of the actual content of student questions and its effective utilization. Relying solely on either skills (row 3) or questions (row 4) is suboptimal, demonstrating their synergistic contribution. Removing both questions and skills (row 5) significantly degrades the model's performance.

The results suggest that the superior performance of SQKT stems from the unique insights provided by student questions, such as their understanding of theoretical concepts and specific struggles, which are not always apparent in code submissions alone. Further, the additional step of explicitly identifying skills from their questions appears to further enhance the clarity of student performance. 

% These results validate the effectiveness of our multi-feature fusion approach, demonstrating that SQKT provides a more accurate representation of students' knowledge states by capturing aspects of learning beyond code submissions.

\paragraph{Impact of Auxiliary Losses}
% \begin{table}[t]
% \centering
% \small
% \setlength{\tabcolsep}{4pt}
% \begin{tabular}{ll c c c}
% \toprule
% Course & Metric(\%) & SQKT & - auxiliary & - triplet \\
% \midrule
% \multirow{3}{*}{Python Introduction} & AUC & \bfseries 92.0 & 91.6 & 90.5 \\
%                                      & ACC & \bfseries 90.6 & 90.0 & 90.4 \\
%                                      & F1  & \bfseries 89.2 & 89.3 & 88.9 \\
% \addlinespace
% \multirow{3}{*}{Python Basic} & AUC & 87.7 & \bfseries 87.9 & 83.7 \\
%                               & ACC & \bfseries 85.6 & 84.3 & 80.0 \\
%                               & F1  & \bfseries 88.2 & 86.9 & 83.9 \\
% \addlinespace
% \multirow{3}{*}{First Python} & AUC & 89.6 & 88.1 & \bfseries 90.7 \\
%                               & ACC & \bfseries 84.3 & 79.5 & 81.9 \\
%                               & F1  & \bfseries 83.9 & 79.9 & 82.0 \\
% \bottomrule
% \end{tabular}
% \caption{Impact of Auxiliary and Triplet Loss Functions}
% \label{table:loss_ablation}
% \end{table}
\begin{table}[t]
\centering
\small
\setlength{\tabcolsep}{0.8pt}
\begin{tabular}{l c c c c c c c c c}
\toprule
 & \multicolumn{3}{c}{Python Intro.} & \multicolumn{3}{c}{First Python} & \multicolumn{3}{c}{Python Basic} \\
\cmidrule(lr){2-4} \cmidrule(lr){5-7} \cmidrule(lr){8-10}
& AUC & ACC & F1 & AUC & ACC & F1 & AUC & ACC & F1 \\
\midrule
SQKT       & \bfseries 92.3 & \bfseries 86.3 & \bfseries 87.3 & \bfseries 93.0 & \bfseries 87.1 &  84.9 & \bfseries 93.3 & \bfseries 88.4 & \bfseries 89.8 \\
- Question & 91.6 & 85.8 & 86.9 & 92.5 & 86.5 & \bfseries 86.9 &  91.9 & 85.7 & 87.7 \\
- Triplet   & 91.3 & 85.8 & 90.1 & 90.1 & 83.6 & 84.9 & 91.5 & 85.8 & 87.7 \\
\bottomrule
\end{tabular}
\caption{Impact of response and triplet loss functions. All values are in percentages.}
\label{table:loss_ablation}
\end{table}

We analyzed the impact of the two auxiliary losses, i.e., question loss (Eq.~\ref{eq:L_T5}) and triplet loss (Eq.~\ref{eq:L_triplet}). 
The results in Table~\ref{table:loss_ablation} validate the importance of these additional objectives in improving performance across diverse programming courses.
The question loss, derived from the task of predicting educator responses to student questions, impacts performance across the three courses, with slight drops in performance when removed. This loss seems to enrich the embedding space by capturing important information in student questions better. 

The triplet loss, designed to unify the embedding space for heterogeneous input features, has stronger impact, making a notable contribution especially for the ``First Python'' course.
The triplet loss ensures effective integration of diverse data sources. 

\begin{figure}[t]
\centering
\includegraphics[width=1\columnwidth]{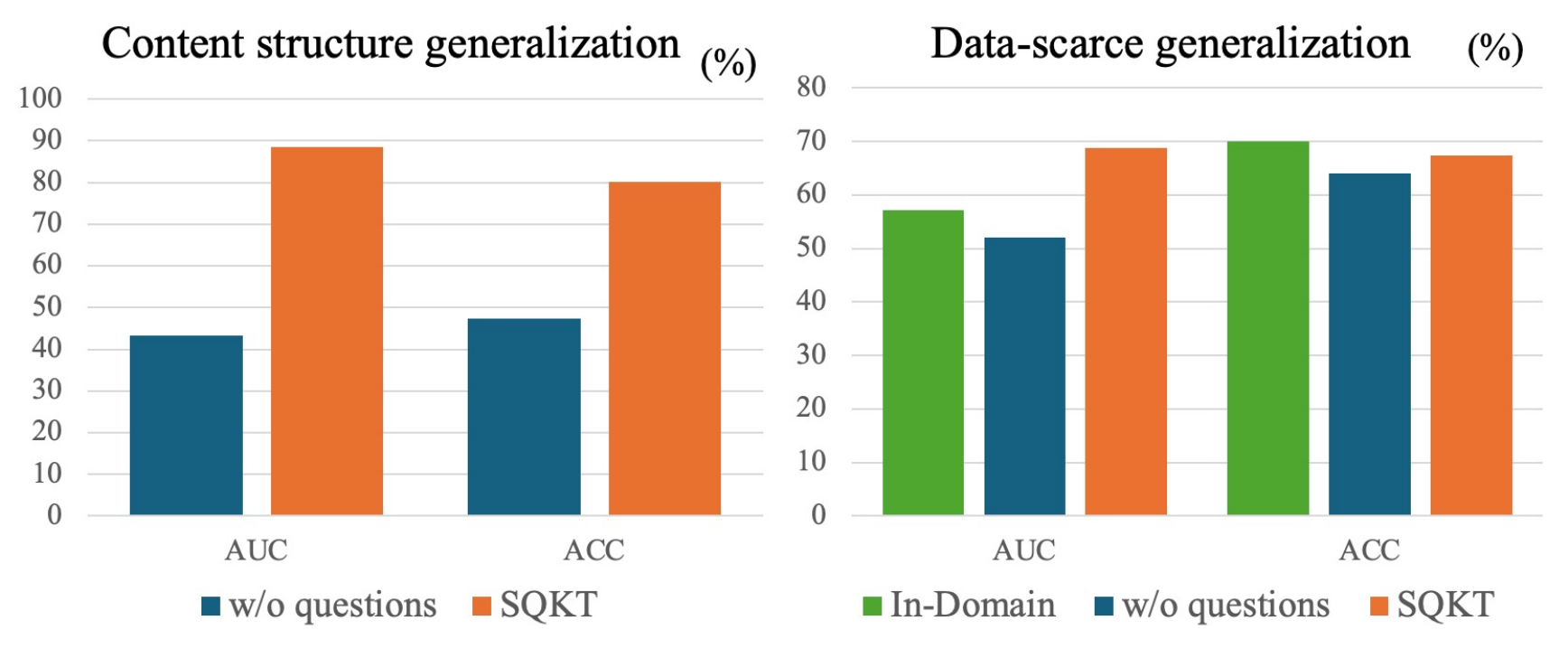} % Reduce the figure size so that it is slightly narrower than the column. Don't use precise values for figure width.This setup will avoid overfull boxes.
\caption{Cross-domain performance.}
\label{cross_fig}
\end{figure}

\subsection{Cross-Domain Results} 
Figure \ref{cross_fig} demonstrates the model's performance across two cross-domain settings, evaluating its ability to generalize to unseen courses.

In the setting of content structure generalization (Figure \ref{cross_fig}, left), we assessed SQKT's ability to transfer knowledge between courses with different levels and content structures. Our full model (orange) showed an absolute 45.3\% improvement in AUC over without using question data (blue).

In the setting of data-scarce generalization (Figure \ref{cross_fig}, right), we trained SQKT on all courses except ``Algorithm'' and tested it on the course to evaluate the model's generalizability to higher difficulty levels and robustness in low-resource environments. Since the ``Algorithm'' course has small data, fine-tuning SQKT directly on the ``Algorithms'' data (green) shows an AUC score close to random. However, our full model (orange) showed a substantial improvement of 11.4\% over the in-domain model (green) and 16.7\% over the cross-domain model incorporating no student questions (blue). 

Both experiments conclude that student questions convey generalizable insights into student performance across different courses and that leveraging them greatly enhances the model's ability to adapt to new courses with varying difficulty levels and limited data.
%underscoring the importance of question data in enhancing robustness in data-scarce settings.

\subsection{Error Analysis} 
To better understand our model, we conducted a detailed analysis focused on question-related mistakes. We randomly sampled 60 mispredictions from the test set, manually analyzed each data point by tagging one or more error types. Table~\ref{table:error_analysis} in Appendix presents a breakdown of these errors, their proportions, examples, and underlying reasons.

Our analysis shows that `Complexity' is the most prevalent issue (55.6\%), often due to code snippets containing mixed language syntax, which challenges the model's parsing capabilities. `Confusion' is the second most common error type (40.7\%), typically occurring when the error in the code is unrelated to the student's question, making it difficult for the model to establish the correct correlation. `Ambiguity' (22.2\%) and `Incompleteness' (29.6\%) also contribute significantly to model errors, emphasizing the need for clear, context-rich questions for accurate predictions. The analysis highlights key areas for improvement in the SQKT model. For example, incorporating more advanced natural language processing techniques to handle multi-lingual input could enhance the model's ability to interpret students' questions more accurately.

\section{Conclusion}
This paper introduces SQKT, a knowledge tracing model in programming education that addresses the unique challenges of predicting students' performance on subsequent problems in coding tasks. By integrating students' questions and auto-extracted skill information, SQKT provides a more comprehensive view of student knowledge than traditional KT models. We demonstrate the effectiveness of SQKT across various programming courses and difficulty levels, consistently outperforming baseline models in both in-domain and cross-domain settings. SQKT shows its ability to capture valuable information about students' programming competencies through their questions.
We expect that our method can contribute to more personalized and effective learning interventions in programming education.
% This allows educators to assign problems that match each student’s level, creating a more tailored and effective learning experience. Furthermore, the model helps identify areas where students may need more support, facilitating a data-driven, personalized approach that aligns with modern educational theories focused on personalization in programming education. However, our error analysis reveals limitations in handling student questions. These limitations highlight areas for future improvement. We hope that our approach will inspire further research in developing more accurate and personalized learning interventions.

\section*{Limitations}
This study has several limitations. First, we did not apply any preprocessing to the input questions prior to analysis. Although this approach more closely mirrors actual classroom conditions, inputs are often noisy. To address this, we implemented a skill extractor system designed to effectively extract information from such noisy inputs. Future research could explore whether introducing filtering or normalization steps might improve model performance. 

Second, the skill extractor system employs a rule-based methodology rather than statistical machine learning techniques. This choice aims to ensure interpretability, offering a clear and explainable mapping between questions and skills. However, adopting machine learning methods could offer significant benefits, such as improved scalability and the ability to adapt to unseen patterns.
Future studies could investigate hybrid approaches that combine the rule-based systems with machine learning models.

\bibliography{custom}

\appendix
\onecolumn
\section{Python Skill Set}

\begin{table}[H]
\small
% \scriptsize
\setlength{\tabcolsep}{2pt}
\begin{tabular}{>{\raggedright\arraybackslash}p{0.18\linewidth} >{\raggedright\arraybackslash}p{0.18\linewidth} >{\raggedright\arraybackslash}p{0.18\linewidth} >{\raggedright\arraybackslash}p{0.18\linewidth} >{\raggedright\arraybackslash}p{0.18\linewidth}}
\toprule
Value & Variable Assign & Keywords & Operators & Operands \\
Type Convertor & input function & print function & Boolean Values & Boolean Expressions \\
Logical Operators & If-Else Statements & For Loops & While Loops & Break Statement \\
Continue Statement & Function Definitions & return Statement & Local, Global Scope & Strings \\
String Slicing & Indexing & Lists & Dictionaries & Import Statement \\
random & time & math & Opening files & Reading files \\
Writing files & Closing files & SyntaxError & NameError & TypeError \\
IndentationError & ValueError & AttributeError & IndexError & KeyError \\
TabError & UnicodeDecodeError & FileNotFoundError & ModuleNotFoundError & ZeroDivisionError \\
UnboundLocalError & ImportError & UnicodeEncodeError & LookupError & ConnectionError \\
RuntimeError &  &  &  &  \\
\bottomrule
\end{tabular}
\caption{Categorized Python concepts and errors}
\label{python_concepts_errors}
\end{table}

\section{Description of dataset and train-validation-test split}
\label{appendix:dataset}
\begin{table}[H]
\centering
\small
\begin{tabular}{l p{11cm}}
\toprule
Course Name & Description \\
\midrule
Python Basic& Beginner-level course covering basic python concepts. \\
First Python  & A beginner-friendly course focusing on fundamental Python programming. \\
Algorithm & An intermediate to advanced course on algorithms implemented in Python. \\
\makecell[l]{Python Introduction} & A comprehensive course covering basic to intermediate python concepts. \\
\bottomrule
\end{tabular}
\caption{Description of dataset}
\label{table:description}
\end{table}

\begin{table}[H]
\centering
\small
\begin{tabular}{l c c c c}
\toprule
Course Name & Type & Train & Validation & Test \\
\midrule
\multirow{2}{*}{Python Basic} & \# of students & 128 & 16 & 16 \\
                              & \# of problems & 2,665 & 362 & 412 \\
\midrule
\multirow{2}{*}{First Python} & \# of students & 6512 & 814 & 815 \\
                              & \# of problems & 153,015 & 19,140 & 11,302 \\
\midrule
\multirow{2}{*}{Algorithm} & \# of students & 61 & 8 & 8 \\
                           & \# of problems & 273 & 52 & 5 \\
\midrule
\multirow{2}{*}{Python Introduction} & \# of students & 873 & 109 & 110 \\
                                     & \# of problemss & 65,372 & 4,690 & 4,135 \\

\bottomrule
\end{tabular}
\caption{Statistics for dataset splits}
\label{table:split_data}
\end{table}

\clearpage

\section{Error Analysis}
\FloatBarrier
\begin{table}[H]
\centering
\begin{tabular*}{\textwidth}{@{\extracolsep{\fill}}p{0.15\textwidth} r p{0.35\textwidth}p{0.25\textwidth}}
\toprule
Error Type & Occurrence & Questions & Reason \\
\midrule
Ambiguity & 22.2\% & \raggedright Traceback (most recent call last): File "main.py", line 4, in $\langle$module$\rangle$ pr NameError: name 'pr' is not defined\\
\textit{"please show me the correct answer."}& The meaning of the question lacks context, allowing for various interpretations  \\
\addlinespace
Confusion &{40.7\%} & \raggedright  
Traceback (most recent call last): File "main.py", line 5, in
$\langle$module$\rangle$ for key in len(int(data)) TypeError: int() argument must be a string, a bytes-like object or a number, not 'dict' \\
\textit{"If len(scores), does it include up to scores? \\
If len(scores), shouldn't it be scores+1 to include the last score?"} & When the error message is unrelated to the student's question, making it difficult to easily determine their correlation, causing the model to be confused about what to focus on within the question \\
\addlinespace
Incompleteness & 29.6\% & \raggedright \textit{"Please provide the answer}\textit{"} & The question is too simple and lacks necessary information\\
\addlinespace
Complexity & 55.6\% & \raggedright File "main.py", line 5 \\
init(self, "\textbf{\textit{sweet potato(Korean)}}"): \\
SyntaxError: invalid syntax \\
\textit{"please explain this error"}& The question contains a code snippet with a mix of Korean characters and English syntax. This combination of different character sets and languages introduces additional complexity\\
\bottomrule
\end{tabular*}
\caption{Detailed error analysis of SQKT}
\label{table:error_analysis}
\end{table}

% \begin{table*}[t]
% \centering
% \small
% \begin{tabular}{c c l}
% \toprule
% Category & Data & Example  \\
% \midrule
% \multirow{5}{*}{\makecell{Alice the Rabbit's \\ Math Homework}} & \raggedright Problem Description & Write a program that takes a natural number as input and outputs the difference between the square of the sum and the sum of the squares for numbers from $1$ to the given input. Specifically, compute and print:

% \[
% \text{Difference} = (\text{Sum of numbers})^2 - (\text{Sum of squares of numbers})
% \]

% \\
% & First Python & 167,453 \\
% & Algorithm & 453 \\
% & Python Introduction & 58,580 \\
% \bottomrule
% \end{tabular}
% \caption{Summary and key statistics of dataset}
% \label{table:data}
% \end{table*}

% \FloatBarrier
\section{Dataset Example}
\label{appendix:example}
\subsection{Python Basic - 1}
\begin{table}[H]
\centering
\small
\begin{tabular}{c c p{11cm}}
\toprule
Problem & Data & Example \\
\midrule
\multirow{6}{*}{\makecell{Alice \\ the \\ Rabbit's \\ Math \\ Homework}} 
& \makecell{Problem \\ Description} 
& \parbox{11cm}{ Write a program that takes a natural number as input and outputs the difference between the square of the sum and the sum of the squares for numbers from $1$ to the given input. } \\ \cmidrule(l){2-3}
& \makecell{Problem \\ Solution} & \parbox{11cm}{% 
\texttt{
N = int(input()) \newline
i\_square = 0 \newline
i\_list = list(range(1, N + 1)) \newline
for i in i\_list: \newline
\hspace*{1em} i\_square += i**2 \newline
\hspace*{1em} i += 1 \newline
sum\_square = sum(i\_list)**2 \newline
print(sum\_square \texttt{-} i\_square)}} \\ \cmidrule(l){2-3}
& \makecell{Student's  \\ code \\ submission} & \parbox{11cm}{% 
\texttt{summation = 0 \newline
while num \texttt{>} 0: \newline
\hspace*{1em} summation = summation + 1 \newline
\hspace*{1em} num = num - 1 \newline
print(summation)}}  \\ \cmidrule(l){2-3}
& \makecell{Student's \\ question} & \parbox{11cm}{Why does the summation variable not produce the correct summation when printed in the
given code?} \\ \cmidrule(l){2-3}
& \makecell{Skill} & \parbox{11cm}{While-loop, Print function, Operator} \\ \cmidrule(l){2-3} 
& \makecell{Educator's \\ response} & \parbox{11cm}{Since the while loop increments summation by 1 in each iteration, if you input 10, the final value stored in summation will be 10.} \\ 
\bottomrule
\end{tabular}
\caption{Example of Python Basic dataset - 1}
\label{appendix:b}
\end{table}

\FloatBarrier
\subsection{Python Basic -2}
\begin{table}[H]
\centering
\small
\begin{tabular}{c c p{11cm}}
\toprule
Problem & Data & Example  \\
\midrule
\multirow{6}{*}{\makecell{Script \\ Polishing}} 
& \makecell{Problem \\ Description} & \parbox{11cm}{%
\vspace{0.2cm}
The variable `sentence` contains a randomly generated even-length sentence read by the Mad Hatter. \newline
\newline
Prompt the user to input a special character to insert into the middle of the sentence. Then, insert the inputted special character into the middle of the string `sentence` and save the result. \vspace{0.2cm}} \\ \cmidrule(l){2-3}
& \makecell{Problem \\ Solution} & \parbox{11cm}{%
\vspace{0.2cm}
\texttt{sentence = \\ sentence[: len(sentence) // 2] + input() + sentence[len(sentence) // 2 :]}%
\vspace{0.2cm}} \\ \cmidrule(l){2-3}
& \makecell{Student's \\ Code \\ Submission} & \parbox{11cm}{%
\vspace{0.2cm}
\texttt{st\_len1 = sentence[x:]} \newline
\texttt{st\_len2 = sentence[:x]} \newline
\texttt{add\_st = str(input())} \newline
\texttt{sentence = st\_len1 + add\_st + st\_len2} \vspace{0.2cm}} \\ \cmidrule(l){2-3}
& \makecell{Student's \\ Question} & \parbox{11cm}{
\vspace{0.2cm} Can’t it be done using only parentheses? \newline
Is it better to use square brackets for distinction? Square brackets are used for index slicing. \newline
Also, how can I insert a string into the middle of another string? \vspace{0.2cm}} \\ \cmidrule(l){2-3}
& Skill & \parbox{11cm}{String, Operators, Indexing} \\ \cmidrule(l){2-3}\
 & \makecell{Educator's \\ response} & \parbox{11cm}{\vspace{0.2cm} Parentheses do not function the same way. \vspace{0.2cm}}\\
\bottomrule
\end{tabular}
\caption{Example of Python Basic dataset - 2}
\label{appendix:c}
\end{table}

\clearpage
\FloatBarrier
\subsection{Python Introduction - 1}
\begin{table}[H]
\centering
\small
\begin{tabular}{c c c}
\toprule
Problem & Data & Example  \\
\midrule
\multirow{6}{*}{\makecell{Copycat \\ Parrot}} 
& \makecell{Problem \\ Description} & \parbox{11cm}{\vspace{0.2cm}If you’ve entered the code on line 02, click [Run]. Do you see the cursor blinking in the output window? Type anything you want to say in this area, then press [Enter].\vspace{0.2cm}} \\ \cmidrule(l){2-3}
& \makecell{Problem \\ Solution} & \parbox{11cm}{%
\vspace{0.2cm} %
\texttt{var = input() \newline
print('Parrot:', var)} \vspace{0.2cm}} \\ \cmidrule(l){2-3}
& \makecell{Student's \\ Code \\ Submission} & \parbox{11cm}{%
\vspace{0.2cm} %
\texttt{var = raw\_input("Enter a value:") \newline
print('Parrot:', var)} \vspace{0.2cm}} \\ \cmidrule(l){2-3}
& \makecell{Student's \\Question} & \parbox{11cm}{%
\vspace{0.2cm} %
var = input('') \newline
print('Parrot:', var) \newline
What should I enter in the input('') function? \vspace{0.2cm}} \\ \cmidrule(l){2-3}
& Skill & \parbox{11cm}{\vspace{0.2cm} Variable Assign, Operands, input function, print function \vspace{0.2cm}} \\ \cmidrule(l){2-3}
& \makecell{Educator's \\ response} & \parbox{11cm}{%
\vspace{0.2cm} % 
Please enclose the string "Parrot" in quotation marks when entering it. For example, you can input it as follows: \newline
\newline
var = input('Parrot') \vspace{0.2cm}} \\ 
\bottomrule
\end{tabular}
\caption{Example of Python Introduction dataset - 1}
\label{appendix:d}
\end{table}

\clearpage
\FloatBarrier
\subsection{Python Introduction - 2}
\begin{table}[H]
\centering
\small
\begin{tabular}{c c c}
\toprule
Problem & Data & Example  \\
\midrule
\multirow{6}{*}{\makecell{Creating a \\ Mysterious \\ Data \\ Dictionary}} 
& \makecell{Problem \\ Description} & \parbox{11cm}{
\vspace{0.2cm}
Enter your age as a number inside the parentheses. \newline
Enter your name as a string inside the parentheses. \newline
Enter a list containing your age and name inside the parentheses.\vspace{0.2cm}
}  \\ \cmidrule(l){2-3}
& \makecell{Problem \\ Solution} & \parbox{11cm}{%
\vspace{0.2cm} %
\texttt{print(20)} \newline
\texttt{print("Your Name")} \newline
\texttt{print([20, "Your Name"])} \vspace{0.2cm}
} \\ \cmidrule(l){2-3}
& \makecell{Student's \\ Code \\ Submission} & \parbox{11cm}{%
\vspace{0.2cm} %
\texttt{print(17)} \newline
\texttt{print("Your Name")} \newline
\texttt{print[17, James Bond]} \vspace{0.2cm}
} \\ \cmidrule(l){2-3}
& \makecell{Student's \\ Question} & \parbox{11cm}{%
\vspace{0.2cm} %
\texttt{Traceback (most recent call last):} \newline
\hspace*{1em} \texttt{  File "main.py", line 8, in <module>} \newline
\hspace*{1em} \texttt{    print[17, James Bond]} \newline
\texttt{NameError: name 'James Bond' is not defined}
\vspace{0.2cm}}\\ \cmidrule(l){2-3}
& Skill & \parbox{11cm}{Value, NameError ]} \\ \cmidrule(l){2-3}
& \makecell{Educator's \\ response} & \parbox{11cm}{%
\vspace{0.2cm} %
This error indicates that the variable \texttt{James Bond} has not been defined. Variables must be defined before they are used.
To fix this, the line \texttt{print[17, James Bond]} needs to be corrected by defining \texttt{James Bond} as a string. \vspace{0.2cm}}  \\
\bottomrule
\end{tabular}
\caption{Example of Python Introduction dataset - 2}
\label{appendix:e}
\end{table}

\FloatBarrier
\subsection{First Python - 1}
\begin{table}[H]
\centering
\small
\begin{tabular}{c c c}
\toprule
Problem & Data & Example  \\
\midrule
\multirow{6}{*}{\makecell{If~Else \\ Statements}} 
& \makecell{Problem \\ Description} & \parbox{11cm}{%
\vspace{0.2cm} %
Write the if condition on line 4 so that it evaluates to true when the entered password \texttt{password} matches the set password \texttt{answer}. \newline
Run the program and try entering the password \texttt{34566}.
 \vspace{0.2cm}}  \\ \cmidrule(l){2-3}
& \makecell{Problem \\ Solution} & \parbox{11cm}{%
\vspace{0.2cm} %
\texttt{answer = 12345} \newline
\texttt{password = input("Enter the password: ")} \newline
\texttt{if password == answer:} {print("Password OK!")} \newline
\texttt{else:} \texttt{print("Password Not OK!")}
\vspace{0.2cm}}  \\ \cmidrule(l){2-3}
& \makecell{Student's \\ Code \\ Submission} & \parbox{11cm}{%
\vspace{0.2cm} %
\texttt{answer = '12345'} \newline
\texttt{password = input('Enter the password: ')} \newline
\texttt{if answer==password:} \texttt{print('Password OK!')} \newline
\texttt{else:} \texttt{print('Password Not OK!')}
\vspace{0.2cm}} \\ \cmidrule(l){2-3}
& \makecell{Student's \\ Quesion} & \parbox{11cm}{
\vspace{0.2cm} If I input '12345' in the terminal, it shows incorrect. But if I input 12345, it shows correct. Why is that? Can’t I set answer = 12345 instead? If answer = '12345', do I need to type 12345 in the terminal for it to match? \vspace{0.2cm}} \\ \cmidrule(l){2-3}
& Skill & \parbox{11cm}{\vspace{0.2cm} input function, Variable \vspace{0.2cm}} \\ \cmidrule(l){2-3}
& \makecell{Educator's \\ response} & \parbox{11cm}{%
\vspace{0.2cm} %
The symbols \texttt{''} are used to represent a string. \newline
Since input values in the terminal are automatically processed as strings, there is no need to include \texttt{''} when typing input in the terminal.
\vspace{0.2cm}} \\
\bottomrule
\end{tabular}
\caption{Example of First Python dataset - 1}
\label{example:f}
\end{table}

\FloatBarrier
\subsection{First Python - 2}
\begin{table}[H]
\centering
\small
\begin{tabular}{c c c}
\toprule
Problem & Data & Example  \\
\midrule
\multirow{6}{*}{\makecell{Prime \\ Number \\ Finder}} 
& \makecell{Problem \\ Description} & \parbox{11cm}{%
\vspace{0.2cm} %
Write a program to find prime numbers between \texttt{1} and \texttt{N}, where \texttt{N} is an input value.
Run the program and enter \texttt{200} as the value of \texttt{N}.\vspace{0.2cm}
}  \\ \cmidrule(l){2-3}
& \makecell{Problem \\ Solution} & \parbox{11cm}{%
\vspace{0.2cm} %
\texttt{n = int(input("Enter the value of N: "))} \newline
\texttt{for a in range(2, n+1):} \newline
\hspace*{1em} \texttt{prime\_yes = True} \newline
\hspace*{1em} \texttt{for i in range(2, a):} \newline
\hspace*{2em} \texttt{if a \% i == 0:}  \texttt{prime\_yes = False} \texttt{break} \newline
\hspace*{1em} \texttt{if prime\_yes:}\texttt{print(a, end=" ")}
 \vspace{0.2cm}
} \\ \cmidrule(l){2-3}
& \makecell{Student's \\ Code \\ Submission} & \parbox{11cm}{%
\vspace{0.2cm} %
\texttt{n = int(input("Enter the value of N: "))} \newline
\texttt{for a in range(2, n+1):} \newline
\hspace*{1em} \texttt{result = True} \newline
\hspace*{1em} \texttt{for i in range(2, a):} \newline
\hspace*{2em} \texttt{if a \% i == 0:} \texttt{result = False} \newline
\hspace*{2em} \texttt{break} \newline
\hspace*{1em} \texttt{if result = True:} \texttt{print(a, end=" ")}
 \vspace{0.2cm}
} \\ \cmidrule(l){2-3}
& \makecell{Student's \\ Question} & \parbox{11cm}{%
\vspace{0.2cm} %
In this part, doesn't `if result:' mean the same as `if result = True'?  \\
Why does it cause an error when I write `if result = True'?
\vspace{0.2cm}}\\ \cmidrule(l){2-3}
& Skill & \parbox{11cm}{\vspace{0.2cm} If-Else Statements, Boolean Values \vspace{0.2cm}} \\ \cmidrule(l){2-3}
& \makecell{Educator's \\ response} & \parbox{11cm}{%
\vspace{0.2cm} 
You need to use ==. \vspace{0.2cm}}  \\
\bottomrule
\end{tabular}
\caption{Example of First Python dataset - 2}
\label{example:g}
\end{table}

\end{document}